\documentclass[12pt]{article}

\usepackage[pdftex]{graphicx}
\title{{\bf Accelerator Disaster Scenarios, the Unabomber, and Scientific Risks}}

\author{{Joseph I. Kapusta}\footnote {Joseph I. Kapusta received his Ph.D. degree at the University of California at Berkeley in 1978 and has been on the faculty of the School of Physics and Astronomy at the University of Minnesota since 1982.  He is the author of over 150 papers in refereed journals and conference proceedings and of {\it Finite Temperature Field Theory} (Cambridge University Press, 1989; second edition with Charles Gale, 2006).}}

\date{}

\begin{document}

\maketitle

\vspace*{-0.8cm}

\begin{abstract}
The possibility that experiments at high-energy accelerators could create new forms of matter that would ultimately destroy the Earth has been considered several times in the past quarter century.  One consequence of the earliest of these disaster scenarios was that the authors of a 1993 article in {\it Physics Today} who reviewed the experiments that had been carried out at the Bevalac at Lawrence Berkeley Laboratory were placed on the FBI's Unabomber watch list.  Later, concerns that experiments at the Relativistic Heavy Ion Collider at Brookhaven National Laboratory might create mini black holes or nuggets of stable strange quark matter resulted in a flurry of articles in the popular press. I discuss this history, as well as Richard A. Posner's provocative analysis and recommendations on how to deal with such scientific risks. I conclude that better communication between scientists and nonscientists would serve to assuage unreasonable fears and focus attention on truly serious potential threats to humankind.
\end{abstract}


\topskip=27pt
\section*{Introduction}

I was a graduate student at the University of California at Berkeley in 1975, exploring the possibility of doing research in theoretical nuclear physics, when I learned that a state of abnormal nuclear matter might be created in heavy-ion collisions in a new accelerator complex, the Bevalac, which was then just going into operation at Lawrence Berkeley Laboratory (LBL).\footnote{The name has since changed to Lawrence Berkeley National Laboratory (LBNL).}  A small group of physicists had examined the possibility that this new state of nuclear matter might grow by accretion and within a matter of seconds gobble up the entire Earth.

One consequence of this disaster scenario was that two of my friends and colleagues who had published an article in {\it Physics Today} about these experiments at the Bevalac were placed on the Federal Bureau of Investigation's bomb watch list in 1994 for fear that they might be targets of the Unabomber.  Later, related disaster scenarios were examined for experiments that were proposed to be carried out at the Relativistic Heavy Ion Collider (RHIC) at Brookhaven National Laboratory (BNL) beginning in 1999, and other experiments that will soon be carried out at the Large Hadron Collider (LHC) at CERN. 

I discuss these disaster scenarios below, as well as Richard A. Posner's provocative analysis and recommendations for dealing with such potential scientific risks.

\section*{Graduate Research at Berkeley}

I entered graduate school at the University of California at Berkeley in the fall of 1974 knowing that I wanted to be a theoretical physicist.  I passed the preliminary examination in physics on my first attempt one year later and then began talking to faculty members about research possibilities in general relativity, particle theory, and nuclear physics.  No one was working on general relativity either in the Physics Department or at Lawrence Berkeley Laboratory (LBL), but someone was in the Mathematics Department.  I love mathematics (I had majored in both physics and mathematics as an undergraduate at the University of Wisconsin in Madison), but I wanted to do graduate research in a field that had some contact with experiment.  As I discovered, some physicists have a poor image of mathematicians.  I attended an awards ceremony in the Physics Department in the spring of 1976 where a retiring faculty member gave a colloquium on the history of architecture on the Berkeley campus.  Showing a slide of the Mathematics building -- a high, modern concrete structure completely out of character with the rest of campus -- the speaker said: ``And then there is the math building.  It is made up of little cubicles.  In each cubicle resides a mathematician.  The mathematician is not allowed to leave in the evening until he or she proves a theorem ... or at least a lemma."  Not very exciting.

I (figure 1) next spoke with several elementary-particle theorists.
\begin{figure}[h]
 \centering
\includegraphics[width=3.0in]{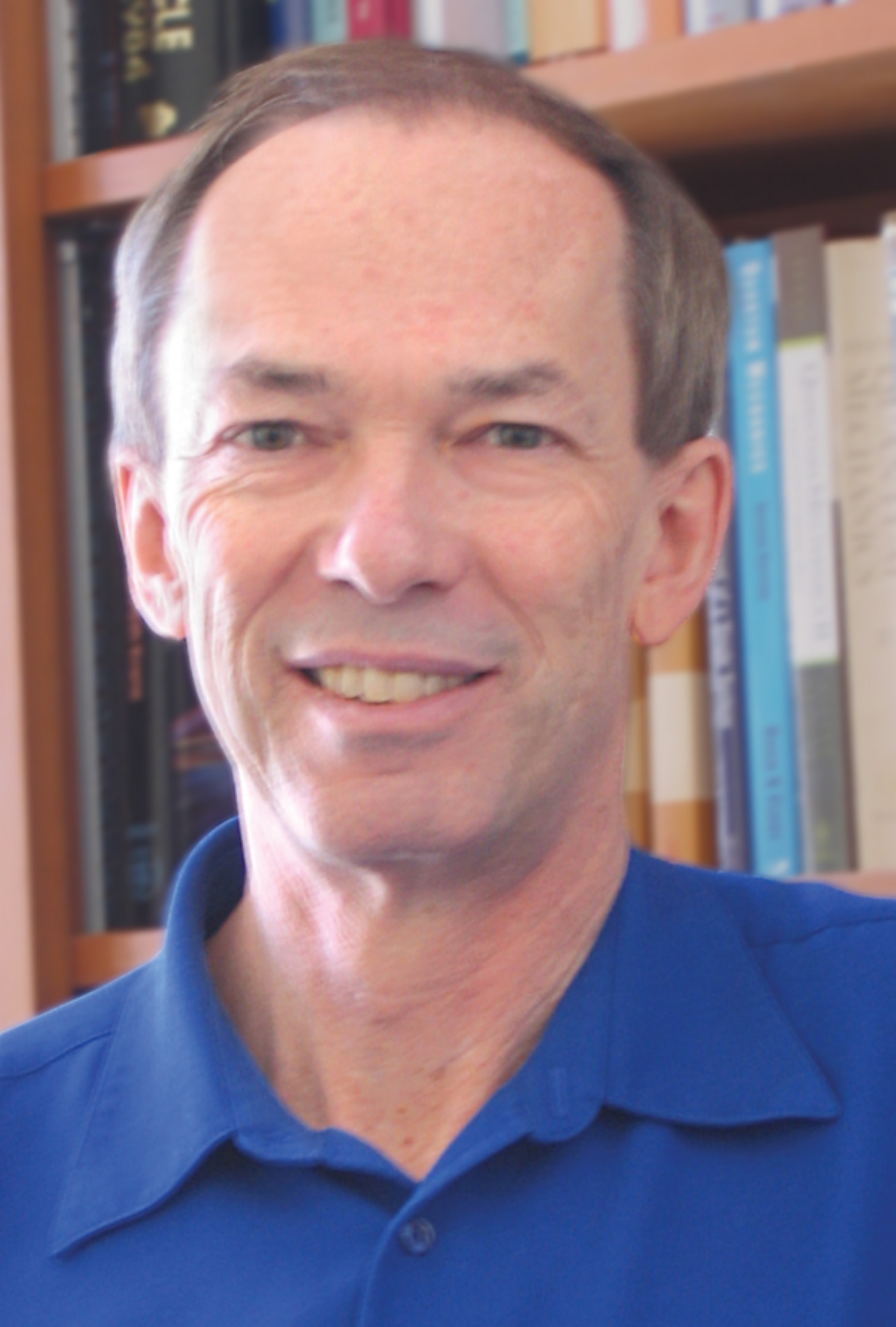}
 \caption{ \footnotesize{Photograph of the author in 2006 by Wendy Tschampl, School of Physics and Astronomy, University of Minnesota.}}
\end{figure}
This was 1975, not long after the work of Sheldon Glashow, Abdus Salam, Steven Weinberg, Gerard 't Hooft, and others on the unification of the weak and electromagnetic interactions, the discovery by David J. Gross and Frank Wilczek and H. David Politzer of asymptotic freedom in Quantum Chromodynamics (QCD) and the rise to power of QCD in strong-interaction physics, and the experimental discovery by Samuel Ting and Burton Richter of the $J/\psi$ particle at Brookhaven National Laboratory (BNL) and at the Stanford Linear Accelerator Center (SLAC).  Every elementary-particle theorist I talked to said: ``Why do you want a Ph.D. in elementary-particle theory?  There aren't any jobs!"  Not very encouraging.

Finally, I walked to the Chemistry Department\footnote{For historical reasons, nuclear-physics research has always been done in the Chemistry Department under Nuclear Chemistry and in the Nuclear Science Division at LBL, not in the Physics Department.} and climbed ``the hill" to LBL to talk with some theoretical nuclear physicists.  I had a very good conversation with Wladek Swiatecki (figure 2),
\begin{figure}[t]
 \centering
\includegraphics[width=3.5in]{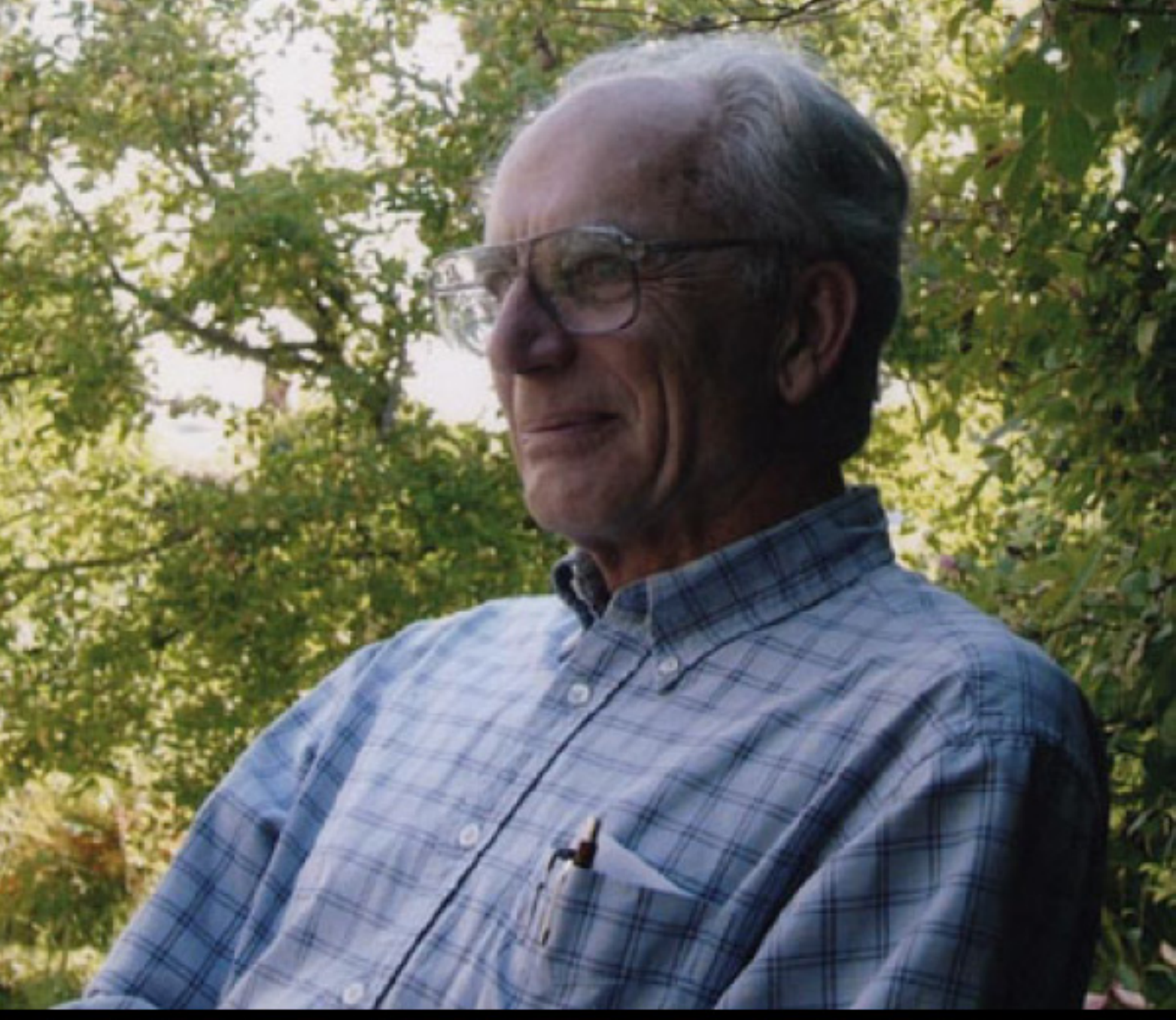}
 \caption{ \footnotesize{Wladek Swiatecki on the occasion of his 80th birthday, summer 2006.  {\it Credit}: Courtesy of Professor Swiatecki.}}
\end{figure}
 who showed me a chart he had made of nuclear radius (which is proportional to the cube root of the atomic weight A) or atomic number Z {\it versus} beam energy (figure 3).
\begin{figure}[h]
 \centering
\includegraphics[width=3.5in]{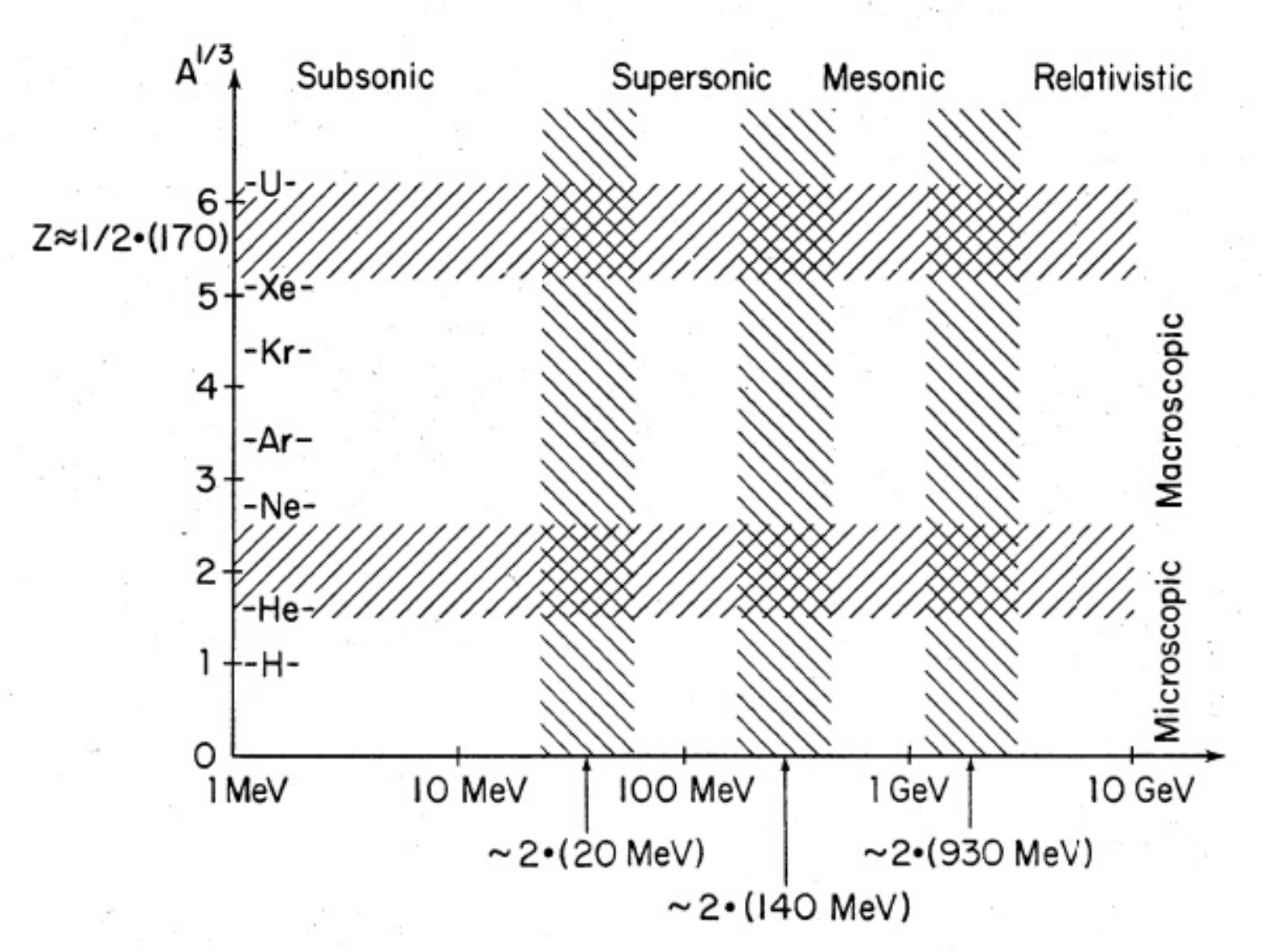}
 \caption{ \footnotesize{Sketch {\it ca}. 1975 of nuclear radius (proportional to the cube root of the atomic weight A) or atomic number Z {\it versus} beam energy showing the various regimes to be explored in particle and nuclear collisions.  {\it Credit}: Courtesy of Wladek Swiatecki.}}
\end{figure}
  He pointed out that the frontier in understanding nuclear structure (represented by the vertical axis) was in the regime of super-heavy nuclei, which depends more upon advances in experiment than in theory.  Further, the goal of high-energy elementary-particle physics (represented by the horizontal axis) was to discover new elementary particles and to deduce their interactions at smaller and smaller distances.  Everything else was virgin territory:  By colliding large, heavy nuclei at higher and higher energies, new and unexplored regimes of high-density nuclear matter would be encountered.  The supersonic regime (collision speed greater than the speed of sound in nuclear matter) would be followed by the mesonic regime (where pion production becomes copious) and then by the relativistic regime (where it is necessary to invoke relativity to describe very dense nuclear matter).

Swiatecki told me, in fact, that Tsung-Dao Lee\footnote{Lee received the Nobel Prize in Physics for 1957 along with Chen Ning Yang for their discovery of  parity nonconservation.} and Gian-Carlo Wick had recently speculated that a new state of nuclear matter might exist at high density where the effective nucleon mass becomes very small, and that this new state, which had to be described by relativistic quantum-field theory, might be metastable or even more stable than ordinary nuclear matter \cite{1}.  This, however, was only speculation, so experiments had to be carried out to search for this new type of nuclear matter -- and now was the time and LBL was the place to do it!  He and others had proposed to connect existing accelerators together to accelerate a beam of large, heavy nuclei to high energy -- to several GeV (giga-electron volts) per beam -- and to smash them into a target of large, heavy nuclei to study the behavior of nuclear matter at high energy.  Their plan was to connect the low-energy heavy-ion accelerator SuperHILAC to the Bevatron\footnote{It was named the Bevatron, because it could accelerate protons to BeV (billion-electron-volt) energies, although the symbol BeV was later replaced by GeV (giga-electron volt) by international agreement; 1 BeV = 1 GeV = $10^9$ eV.} with a transfer line and to call the resulting complex the Bevalac.  The Bevatron (figure 4)\footnote{I first saw this photograph in a textbook that I used as a sophomore in college; see Paul A. Tipler, {\it Foundations of Modern Physics} (New York: Worth Publishers, 1969), p. 49.} could be seen outside Swiatecki's window.  
\begin{figure}[h]
 \centering
\includegraphics[width=4.0in]{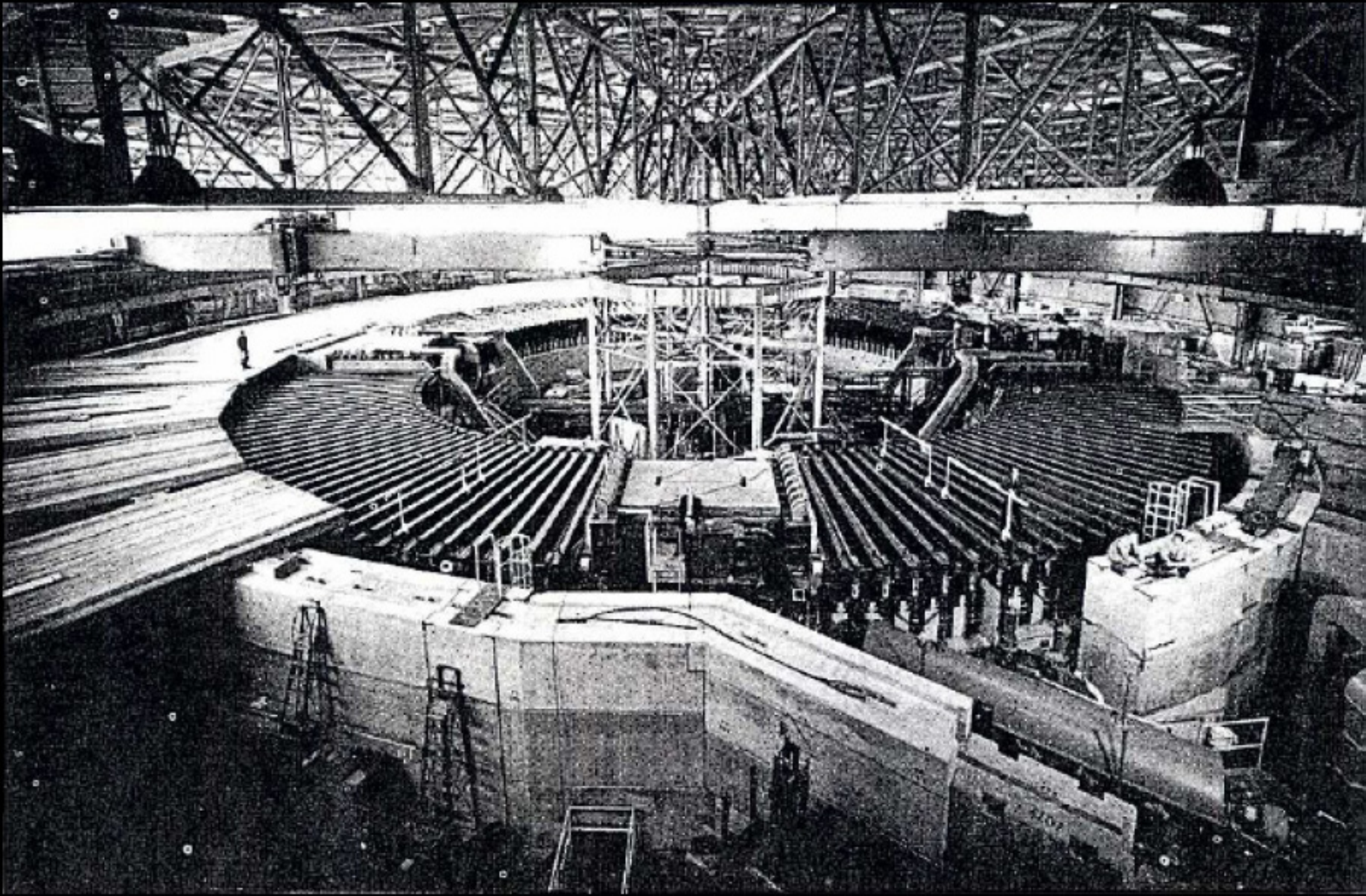}
 \caption{ \footnotesize{The Bevatron accelerator at Lawrence Berkeley Laboratory with which Owen Chamberlain and Emilio Segr\'e discovered the antiproton.  Protons are injected into the linear accelerator shown at the lower right and accelerated to 6.2 GeV.  {\it Credit}: Courtesy of Lawrence Berkeley Laboratory.}}
\end{figure}
It was built in 1954 to discover the antiproton at a beam energy of 6.2 GeV, for which Owen Chamberlain and Emilio Segr\'e shared the Nobel Prize in Physics for 1959.
  
This seemed to me to be a great opportunity for research.  I could get in on the ground floor of a new subfield of physics that would combine aspects of both nuclear and particle physics; I would have to understand and use relativistic quantum-field theory; and I could explore possible applications to the structure of neutron stars and the evolution of the early universe. The Nuclear Science Division at LBL was then at the center of heavy-ion physics. Swiatecki at LBL and Korgut Bardacki in the Physics Department agreed to serve as my advisors.  I completed my Ph.D. degree in 1978, publishing the research I had carried out for my thesis in papers on the nuclear fireball and firestreak models of high-energy nuclear collisions and on QCD at high temperature \cite{2}.

\section*{The Bevalac and Abnormal Nuclear Matter}

The primary purpose in combining the SuperHILAC and the Bevatron to form the Bevalac was to create dense nuclear matter in the laboratory for a brief moment of time.  During 1974-1975 the first beams of carbon and oxygen nuclei were accelerated up to 2.1 GeV per nucleon and smashed into various nuclear targets.   An upgrade was necessary to accelerate uranium nuclei, and in 1981-1982 uranium was accelerated to 1 GeV per nucleon beam energy.  Two major detectors, the Plastic Ball and the Streamer Chamber, were used to measure the spectra of pions, protons, deuterons, helium-3, helium-4, and heavier fragments that were produced.  Ten-to-twenty physicists comprised these collaborations, which was large by experimental nuclear-physics standards at the time but small compared to those in experimental high-energy physics.  The Bevalac was turned off for the last time in 1993, having been eclipsed in energy by the Alternating Gradient Synchrotron (AGS) at Brookhaven National Laboratory (BNL) on Long Island, New York, and by the Super Proton Synchrotron (SPS) at CERN near Geneva, Switzerland.

When the experimental program at the Bevalac began, no one really knew what to expect when nuclear matter was compressed to three-to-four times the density of atomic nuclei.  Since the Fermi momentum increases with density, relativistic quantum-field theory eventually has to be employed to describe the system.  As noted above, in 1974 Lee and Wick suggested that in a limited domain of space a neutral scalar field may acquire an abnormal value (when compared to the rest of the universe), and that this state may be metastable \cite{3}.  If the scalar field has sufficiently strong coupling to nucleons, then their masses would be greatly decreased, leading to a yet-unobserved physical system.  They suggested that this might occur inside a heavy nucleus, but compressing nuclei in heavy-ion collisions was an obvious way to search for this new state of nuclear matter.

The qualitative curves of energy per nucleon ($E/N$) {\it versus} the number of nucleons per unit volume or density of nucleons ($N/V$) shown in figure 5 illustrate these possibilities. 
\begin{figure}
 \centering
\includegraphics[width=5.0in]{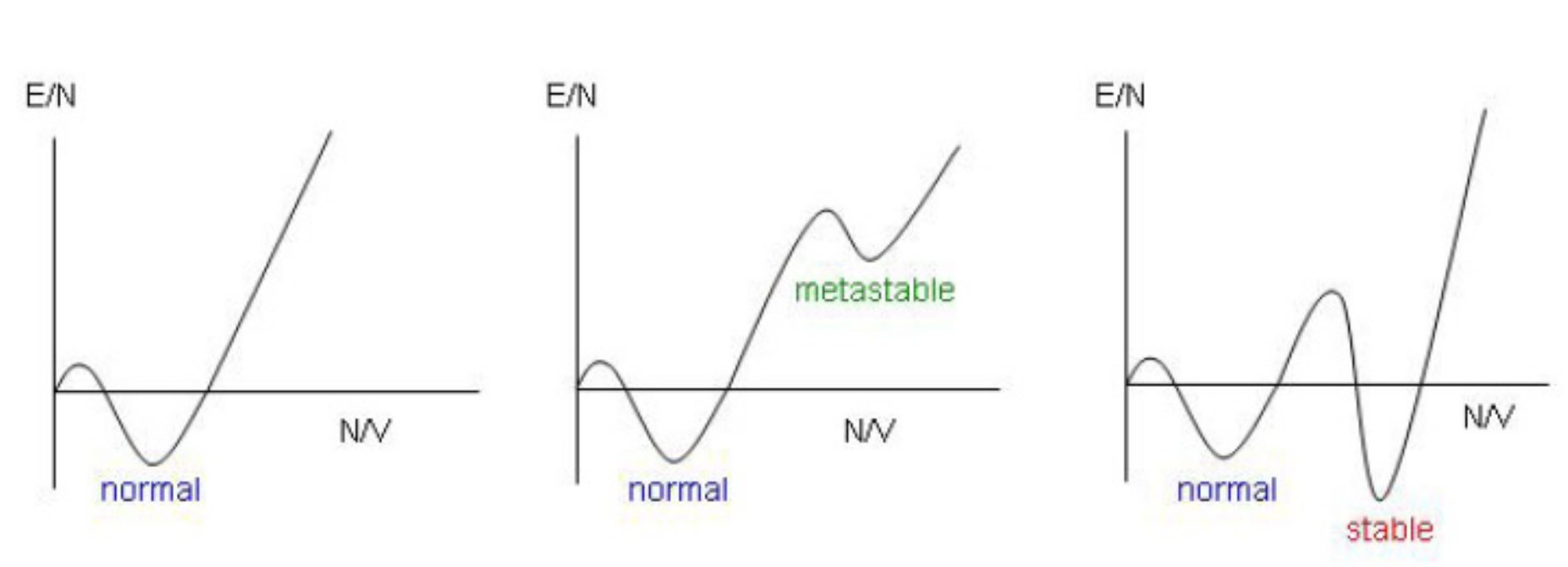}
 \caption{ \footnotesize{Author's sketches of possible nuclear equations of state showing the energy per nucleon $(E/N)$ {\it versus} the number of nucleons per unit volume or density of nucleons $(N/V)$. The curve in the middle and the one on the right illustrate the potential existence of metastable states and ones that are lower in energy than normal nuclei.}}
\end{figure}
The curve on the left, which has a global minimum at the position of ordinary atomic nuclei, illustrates the usual picture. The curve in the middle shows a metastable ``Lee-Wick abnormal state" at some density above the density in atomic nuclei; this state would eventually decay to the lower-energy state.  The curve on the right illustrates an extreme case in which the ``Lee-Wick abnormal matter" lies lower in energy than normal nuclear matter; in this case, ordinary nuclei would eventually decay into this new state of nuclear matter.  Our knowledge about high-density nuclear matter was so poor at this time that no one could rule out these last two possibilities.

Lee and Wick actually were not the first to publish such a speculation:  In 1971 Arnold Bodmer suggested on the basis of quark models and soft interactions between nucleons that collapsed nuclei might be formed \cite{4}.  He called the abnormal states shown in figure 5 isomers in analogy to molecular isomeric states,\footnote{Bodmer thanks Paul J. Ellis for pointing out the analogy to molecular isomers.  Ellis became my friend and collaborator when I joined the Minnesota faculty in 1982; he died suddenly of a heart attack in 2005 as I was preparing an early Powerpoint version of this paper for a talk.}
but they soon came to be called ``density isomers."  For whatever reason, however, Lee and Wick, rather than Bodmer, are usually cited as the originators of the concept of ``abnormal" or ``isomeric" nuclear states.

No one had a clear idea about how the formation of such new abnormal or isomeric states of nuclear matter could be identified in heavy-ion collisions at the Bevalac.  Some said, with tongue-in-cheek, that: ``Heavy-ion collisions will compress the nuclei to such a degree that abnormal nuclear matter will be formed in the core of the compressed nuclei.  This abnormal nuclear matter, being more stable than ordinary matter, will accrete stuff around it and grow to visible size.  Being so massive it will drop to the floor of the experimental hall where one can weigh it and measure its radius, thereby determining its density!"  Such an object, however, would be denser than ordinary nuclear matter ($2 \times 10^{14}$ grams per cubic centimeter) and hence cannot be supported by steel or concrete and would fall to the center of the Earth!  Further, what would prevent it from growing larger and larger until it would occupy the entire Earth?  Simple estimates suggested that this could occur in a matter of seconds -- and if it did no physicist would be around to be blamed for it!  Moreover, it guaranteed that no physicist would ever win a Nobel Prize for the discovery of stable abnormal nuclear matter, since either this new state of nuclear matter does not exist, or the world would end before the Prize could be awarded.  No one took all of this too seriously, and experiments with colliding beams of light and intermediate-mass nuclei proceeded apace.

\section*{The Unabomber}

The case of the Unabomber was one of the most difficult ones ever faced by the U.S. Federal Bureau of Investigation (FBI) \cite{5}.  Between 1978 and 1995, the Unabomber sent bombs through the U.S. mail that detonated when the package was opened, killing three people and maiming or injuring twenty-three more.  The perpetrator was called the Unabomber because he initially targeted scientists, engineers, or technicians working at universities and airlines (UNiversityAirlineBOMBER).  Despite enormous efforts in analyzing letters and other clues, the FBI failed to identify the Unabomber.  Then, in 1995, The {\it Washington Post} and {\it The New York Times} published an eight-page manifesto written by the Unabomber after obtaining a promise that by doing so the bombings would stop \cite{6}.  The Unabomber apparently believed that certain applications of science, engineering, and technology were highly detrimental to human society and had to be stopped; this end justified his means.

One man suspected that his brother was the Unabomber after he noticed a strong similarity in the style of writing between the Unabomber's manifesto and the letters he had received from his brother.  He transmitted his suspicion to the FBI under the condition that if his brother indeed was the Unabomber and was found guilty of his crimes, he would not receive the death penalty.  On April 3, 1996, Theodore (Ted) J. Kaczynski was arrested at his shack near Lincoln, Montana (figure 6), which he had constructed himself and had lived in until then.
\begin{figure}[h]
 \centering
\includegraphics[width=4.0in]{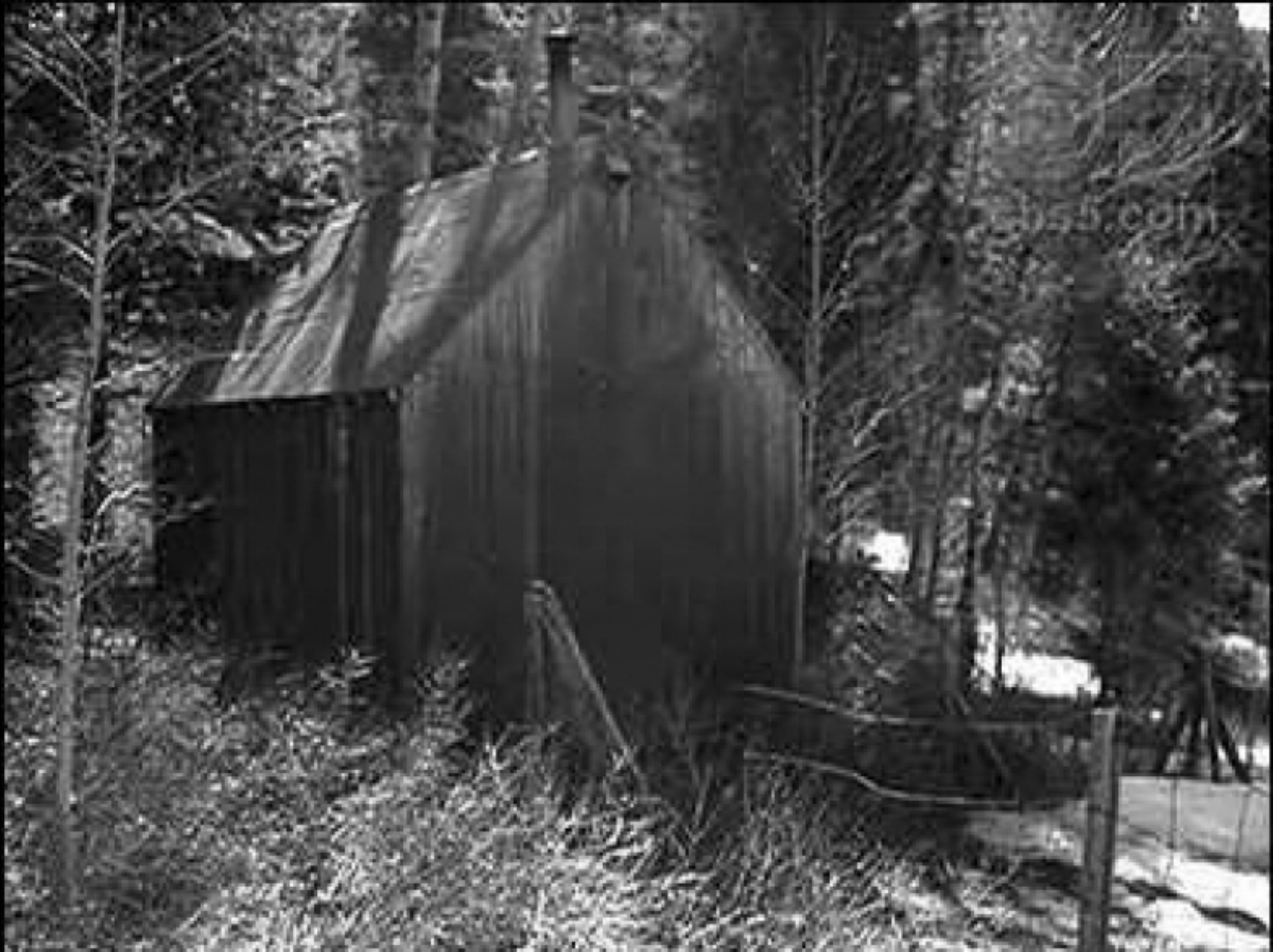}
 \caption{\footnotesize{The shack near Lincoln, Montana, in which the Unabomber, Ted Kaczynski, lived for many years.
{\it Source}:  KPIX TV CBS5, San Francisco; website 
http://cbs5.com/slideshows/unabom.unabomber.exclusive.20.433402.html?rid=8}}
\end{figure}

Kaczynski had earned a B.A. degree in mathematics from Harvard University in 1962 and a Ph.D. degree in mathematics from the University of Michigan in Ann Arbor in 1967, specializing in geometric function theory, a branch of complex analysis.  That fall he was appointed to an assistant professorship in the Mathematics Department at the University of California at Berkeley, from which he resigned without explanation in 1969.  Calvin Moore, Vice Chairman of the Mathematics Department in 1968, said that ``I think he could have advanced along the lines and been a senior member of the faculty." \cite{7}  Kaczynski is now serving a life sentence without the possibility of parole at the Supermax prison in Colorado.

The FBI had placed my good friends and colleagues Subal Das Gupta, Professor of Physics at McGill University in Montreal, and Gary Westfall, Professor of Physics at Michigan State University in East Lansing, on its bomb watch list about a year before Kaczynski's arrest.  Westfall allowed the FBI to search his mail for bombs hidden in packages until a month after Kaczynski's arrest; nothing was ever found.  Das Gupta (figure 7), 
\begin{figure}[h]
 \centering
\includegraphics[width=3.5in]{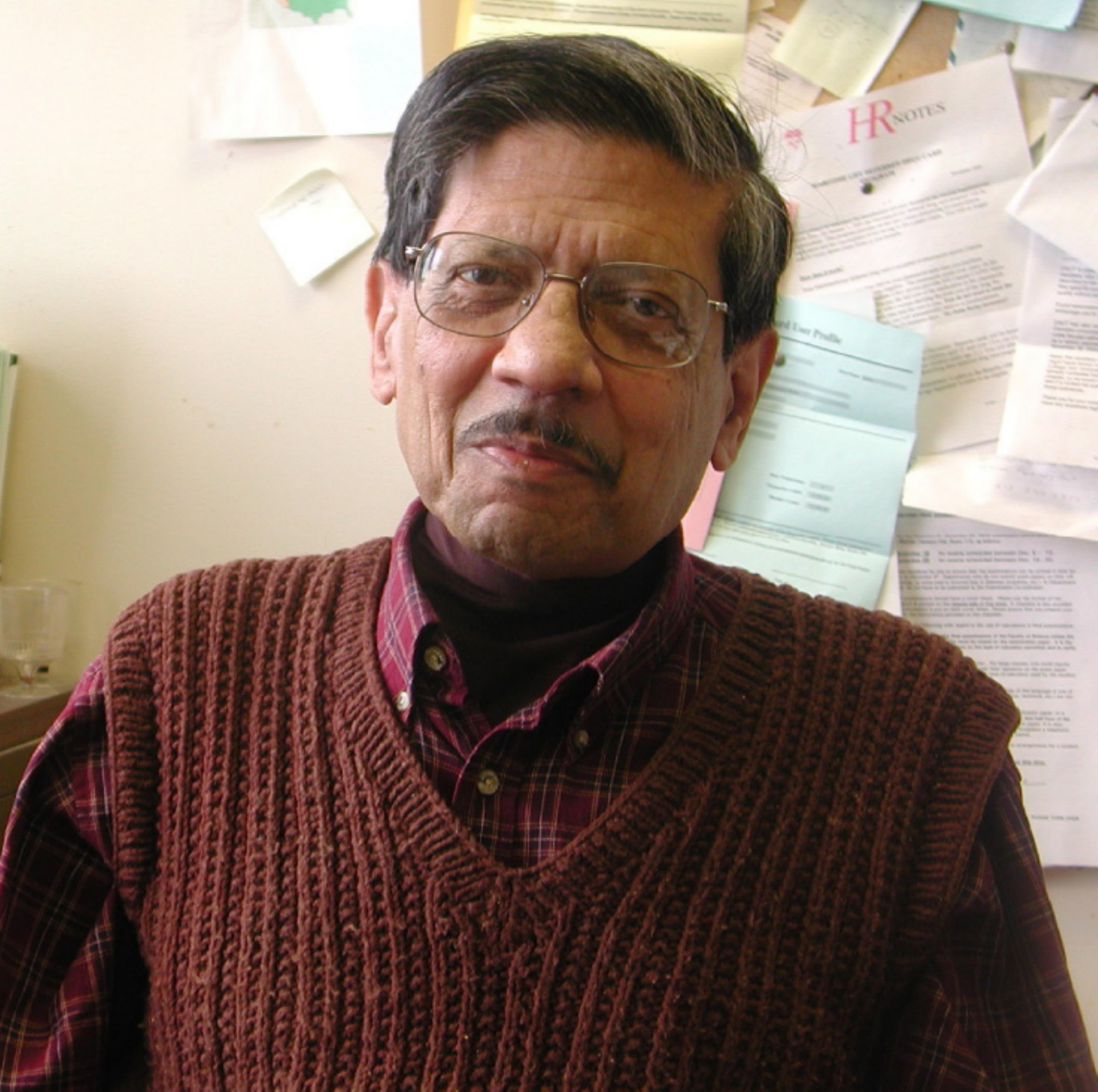}
 \caption{ \footnotesize{Subal Das Gupta, McGill University, Montreal.  {\it Credit}: Courtesy of Professor Das Gupta.}}
\end{figure}
as a Canadian citizen, could refuse to allow the FBI to search his mail, which he did.  I asked him why he did, and he replied: ``I trust the Canadian postal system, the McGill University postal system, and I trust that my secretary would examine any package carefully before she gave it to me."  (He was Chair of the Physics Department at this time.)  Nothing was ever found.  Why then did the FBI place these two physicists on its bomb watch list?

Just before the Bevalac was to be turned off in 1993, I thought that the physics community would be well served if an article were published in {\it Physics Today} that would summarize what had been learned at the accelerator about dense nuclear matter.  I suggested this to Gloria Lubkin, then Editor of {\it Physics Today}; she agreed and asked me to recommend authors for it.  I recommended Westfall, an experimentalist, and Das Gupta, a theorist, both of whom had been involved with the Bevalac since the late 1970s.  After they submitted their manuscript, Lubkin asked me to review it.\footnote{I do not know if my review was anonymous or not, but following standard practice my name was not listed in the acknowledgements.}  I strongly supported its publication \cite{8}. They had pretty well pinned down the nuclear equation of state up to about two-to-three times the normal nuclear-matter density, as well as the momentum dependence of the nuclear-optical potential (the energy of a nucleon as a function of its momentum when passing through nuclear matter of a given density).  I suggested, however, that they should say something about the motivation to join the SuperHILAC to the Bevatron to form the Bevalac, in particular, something  about Lee and Wick's speculation about abnormal nuclear matter.  I drafted a paragraph on this and sent it to Lubkin.  To my surprise and satisfaction, Das Gupta and Westfall thanked me ``for providing the impetus for writing this article," \cite{9} and they incorporated words from my draft paragraph almost unchanged, namely, writing that: ``Meetings were held behind closed doors to decide whether or not the proposed experiments should be aborted."  ``Experiments were eventually performed, and fortunately no such disaster has yet occurred." \cite{10}

The committee that had met behind closed doors included and reported to Bernard Harvey, Associate Director of LBL's Nuclear Science Division; it is dated May 14, 1979, and I provide a transcription of it in the Appendix.  The committee thus met about five years after the first experiments with light ions had begun at the Bevelac, but about two years prior to its upgrade to accelerate heavy ions like uranium.  Thus, there apparently was little concern that colliding light ions would lead to abnormal nuclear matter, but considerable concern that colliding heavy ions might.  In any case, based upon this one-page report the upgrade of the Bevalac was completed and heavy-ion experiments were carried out with it.  No one seriously believed that a disaster of the type imagined could ever occur, given that QCD is the relevant theory of the strong interactions and that high-density nuclear matter should not be described as such, but as quark matter.  In addition, the committee gave observational arguments for its belief.  Nevertheless, this astonishingly brief report was never widely circulated among physicists.  Indeed, my request to the LBNL Director's Office for a copy of it was acknowledged conscientiously, but their search came up empty: The LBNL Director's Office has no official record of it.

The FBI thus placed Das Gupta and Westfall on its bomb watch list, because the FBI thought they might be targets of the Unabomber, since they had written about, and apparently had participated in experiments that might have destroyed the human race.  With the written permission of Das Gupta and Westfall, I requested information from the FBI records under the Freedom of Information Act concerning this issue.  Thankfully, no record of their names were found among any of Kaczynski's papers.

\section*{RHIC, Strangelets, and Black Holes}

The Relativistic Heavy Ion Collider (RHIC) was built at Brookhaven National Laboratory (BNL) specifically to create quark-gluon plasma, a new state of matter representative of the state of the universe when it was less than one microsecond old and temperatures were greater than two trillion Kelvin.  QCD predicts this new state of matter unambiguously because of asymptotic freedom, a key feature of the theory that results in forces becoming weaker and weaker at shorter and shorter distances.  Intuitively, quark-gluon plasma is formed when the density of nucleons, pions, and other hadrons observed in high-energy physics experiments becomes so great that they overlap geometrically; the quarks and gluons then do not know to which hadron they belong and therefore are no longer confined to regions of space characteristic of hadronic sizes, about one femtometer ($10^{-15}$ meter).  The gas of quasi-free quarks and gluons is called plasma, because these entities carry color charge, a nonabelian version of ordinary electric charge.

Like the Bevalac, RHIC owes its heritage to high-energy particle physics.  In the late 1970s and early 1980s the U.S. physics community made plans to construct a machine at BNL to collide counter-rotating beams of protons at energies of 200 GeV per proton in the center-of-momentum frame.  It was called the Colliding Beam Accelerator (CBA) or ISABELLE, but was canceled in 1983 for a variety of scientific, technical, and financial reasons \cite{11}.  The underground circular tunnel to house the accelerator, however, had already been dug by that time,\footnote{The particle-physics community moved on to begin construction of an even bigger machine in Texas, the Superconducting Super Collider (SSC).  The U.S. Congress canceled this project in 1993 after spending \$2 billion of a projected cost of \$10 billion. See Michael Riordan, ``The Demise of the Superconducting Super Collider," {\it Physics in Perspective} {\bf 2} (2000), 411-425.  Congress later ordered the partially dug tunnel for the SSC to be filled in to prevent a resurrection of the project at some time in the future.} so the nuclear-physics community recognized that an opportunity now existed to build RHIC inside the vacant tunnel with the goal of colliding nuclei as massive as gold at energies up to 100 GeV per nucleon per beam.  Theorists estimated that this should be sufficient to create quark-gluon plasma with an energy density of about one hundred times that of ordinary nuclear matter for a very brief moment of time.  RHIC was accorded the highest priority in the Long Range Plan for Nuclear Science, and the U.S. Department of Energy and Congress approved funding for it. Its construction was completed on budget in 1999 (figure 8).
\begin{figure}[h]
 \centering
\includegraphics[width=5.0in]{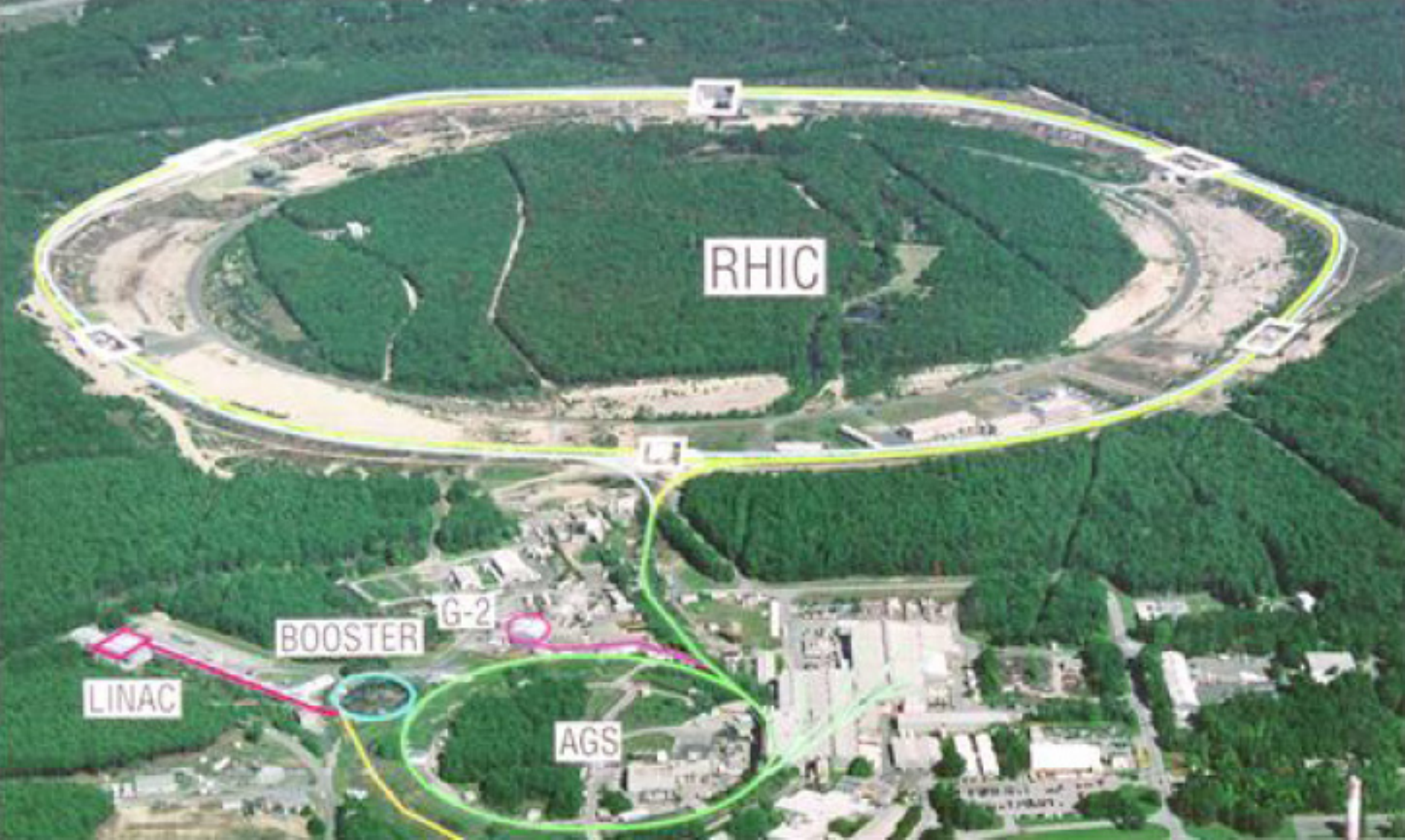}
 \caption{ \footnotesize{The Relativistic Heavy Ion Collider (RHIC) complex at Brookhaven National Laboratory (BNL). The circumference of its ring is 3900 meters.  Gold ions are produced in a Tandem Van de Graaff (not shown) and sent via a transfer line to a Booster to increase their energy.  From there they are sent to the AGS (Alternating Gradient Synchrotron) where their energy is increased further.  Finally they are sent to RHIC, half going in a clockwise direction and half going in a counter-clockwise direction, and accelerated to their final energy.   The LINAC (Linear Accelerator) produces protons for use in the AGS and RHIC.  {\it Credit}: Courtesy of Brookhaven National Laboratory.}}
\end{figure}

Just before RHIC was scheduled to get its first beam, an article by Madhursee Mukerjee entitled ``A Little Big Bang" appeared in {\it Scientific American} in March 1999 \cite{12}.  Mukerjee described the physics that would be investigated at RHIC.  Some of her readers, however, were alarmed by the idea that RHIC would create matter that had not been seen since the Big Bang 13.7 billion years ago.  Thus, Michael Cogill of Coquitlan, British Columbia, wrote: ``I am concerned that physicists are boldly going where it may be unsafe to go... What if they somehow alter the underlying nature of things such that it cannot be restored?" \cite{13}  Walter L. Wagner reacted more specifically by e-mail:  ``My calculations indicate that the Brookhaven collider does not obtain sufficient energies to produce a mini black hole; however, my calculations might be wrong.... Is the Brookhaven collider for certain below the threshold?" \cite{14}

Frank Wilczek of the Institute for Advanced Study in Princeton replied to Cogill and Wagner. He discounted as ``incredible" the scenario that mini black holes would be created at RHIC, but at the same time allowed that ``there is a speculative but quite respectable possibility that subatomic chunks of a new stable form of matter called strangelets might be produced.... But strangelets, if they exist at all, are not aggressive, and they will start out very, very, small." \cite{15}  I doubt that these words were very comforting to a concerned nonscientist, even though Wilczek added that ``here again a doomsday scenario is not plausible."

The {\it Sunday Times} (London) featured an article about RHIC on July 18, 1999, under the provocative banner: ``The final experiment?" \cite{16}  It discussed two possible routes to world destruction.  One involved the production of mini black holes that would grow by accretion until they ultimately would gobble up the Earth.  The other involved the production of strangelets, objects that comprise approximately equal numbers of up, down, and strange quarks, as opposed to protons and neutrons, which comprise up and down quarks only.  Strange quarks are heavier than up and down quarks, so they normally would not be present in ordinary nuclear matter.  At sufficiently high density, however, some up and down quarks might be transformed into strange quarks to lower the Fermi energy, just as having approximately equal numbers of up and down quarks lower the Fermi energy in nuclei, a possibility that Edward Witten and Edward Farhi and Robert Jaffe advanced in 1984 \cite{17}.  Now, if strangelets were formed and were stable, they could gobble up the Earth just like the Lee-Wick abnormal nuclear matter or mini black holes could. To make strangelets, there has to be many up, down, and strange quarks in a small region of space. Did these conditions exist at RHIC?
  
The article in {\it The Sunday Times} was written by its Senior Editor, Jonathan Leake, under the title, ``Big Bang machine could destroy Earth." Leake referred to the first injection of beams into RHIC's circular rings in July 1999 as a ``test-firing" -- as if RHIC's scientists and engineers were testing a weapon instead of a research accelerator.  Leake's article, however, was not inflammatory and did not mention anything that had not been discussed earlier in the {\it Physical Review} or {\it Scientific American}.  In his concluding paragraph he quoted John Nelson, Professor of Physics at the University of Birmingham, as saying: ``The big question is whether the planet will disappear in the twinkling of an eye.  It is astonishingly unlikely that there is any risk 
-- but I could not prove it." \cite{18} These words, like Wilczek's reply to Cogill and Wagner, seems to me to be not very comforting to a concerned nonscientist.
  
There is a postscript to this story. I gave an after-dinner talk on this topic onboard a large boat cruising up and down the blue Danube during the Quark Matter 2005 conference in Budapest, Hungary.  I showed a picture of Leake's article and mentioned his quote by John Nelson.  I jokingly said that I did not know Nelson, but his comment did not seem to me to be very comforting to a layperson.  From far away, near the bow, I heard some commotion and then a loud voice: ``I'm John Nelson -- and I was misquoted!"  A chorus of laughter erupted from bow to stern.

Leake also mentioned in his article that John Marburger III, Director of BNL (and future presidential science advisor), had appointed a committee to investigate whether something ``could go disastrously wrong" during the running of RHIC.  Its members, Wit Busza, Robert Jaffe, Jack Sandweiss, and Frank Wilczek, adapted and revised their report and published it under the title, ``Review of speculative `disaster scenarios' at RHIC" in the {\it Reviews of Modern Physics} \cite{19}.  The main points they made were as follows:
\begin{itemize}
\item A mini black hole with the energy of two colliding gold nuclei at RHIC would have a Schwarzschild radius of $10^{-34}$ femtometer, whereas the radius of a gold nucleus is 7 femtometers.  The probability of a fluctuation that would concentrate all of the energy of two colliding gold nuclei into such a small volume of space occupied by a mini black hole is unimaginably small.
\item An ideal strangelet would have equal numbers of up, down, and strange quarks, represented by the symbol (uds)$^{\textstyle A}$ for an object of baryon or atomic number A, making it electrically neutral.  The strange quark, however, is heavier than the up and down quarks, implying that a strangelet would be slightly depleted of strange quarks, making it positively charged.  It therefore would be repelled by ordinary nuclei, which would inhibit fusion and thus the subsequent conversion into a larger strangelet from ordinary matter.
\item There was a dedicated experiment to search for strangelets at the AGS at BNL using beams of gold nuclei with an energy of 11 GeV per nucleon on fixed targets.  This effort, led by Sandweiss, did not find a single strangelet candidate \cite{20}.  A corresponding search at the SPS at CERN also failed to find strangelets \cite{21}.
\item There is a component of iron in very high-energy cosmic rays, which have been bombarding the moon for billions of years ... and the moon is still there.
\end{itemize}

It seemed to me at the time that Marburger was being very cautious and did the right thing in appointing this committee, whose analyses made good reading but were not particularly startling to the physics community.

\section*{Posner's Proposals}

Richard A. Posner published his thought-provoking book, {\it Catastrophe: Risk and Response}, in 2004 \cite{22}, and several reviews of it appeared thereafter, including one by Kenneth R. Foster in {\it Science} magazine on February 25, 2005, and one by Brian H. Nordstrom in the Newsletter of the Forum on Physics and Society of the American Physical Society in January 2006 \cite{23}.   As the title of his book suggests, Posner raises the questions of how much risk of a major catastrophe we are willing to accept to obtain a certain benefit, and what society's response should be to such risks.  RHIC figures prominently in his book.

Posner was a distinguished Professor of Law at the University of Chicago from 1969 to 1981, after which he was appointed as a Judge on the U.S. Court of Appeals for the Seventh Circuit and served as Chief Judge from 1993 to 2000.  He has written more than thirty books and has served as editor or as a member of the editorial boards of various legal journals and law reviews.  Some of his concerns regarding the possibility that experiments at RHIC might cause the end of the world are as follows:
\begin{itemize}
\item He is concerned that the members of Marburger's RHIC assessment committee (Busza, Jaffe, Sandweiss, Wilczek) were not disinterested parties but had strong motivations to insure that the experiments would be carried out.
\item He cites Italian physicist Francesco Calogero, who suggested in 2000 that BNL should engage two teams, ``a `blue team' trying  to make an `objective' assessment, and a 'red team' (acting as devil's advocates) specifically charged with making a genuine effort at proving that the experiments are indeed dangerous...." \cite{24}
\item He also cites Arnon Dar, Alvaro De Rujula, and Ulrich Heinz, who concluded on the basis of astrophysical considerations that the annual probability that strangelets produced at RHIC will destroy the Earth is less than 1 in 5 million, which compared to other potential disasters may be an unacceptably high probability \cite{25}.
\end{itemize}
 
Posner concludes: ``Congress should consider enacting a law that would require all scientific research projects in specified areas, such as nanotechnology and experimental high-energy physics, to be reviewed by a {\it federal 
catastrophic-risks assessment board and forbidden} if the board found that the project would create an undue risk to human survival [my italics]." \cite{26}
  
Posner's perspective is different from that of most scientists, and is a legitimate one.  Take, for example, his first two points above.  Marburger's RHIC assessment committee included two experimentalists, Busza from the Massachusetts Institute of Technology (MIT), and Sandweiss from Yale University, who to my knowledge have never been involved in the same experimental collaboration.  It also included two theorists, Jaffe and Wilczek,\footnote{Posner refers to Wilczek, who meanwhile moved from the Institute for Advanced Study to MIT, as an experimentalist in this context for reasons that he gives in his book.} both from MIT, who to my knowledge had never collaborated on any research paper prior to the publication of the committee's report, and on only a few thereafter.  A physicist thus would be inclined to conclude that the committee produced a fair and unbiased report on the potential disasters that might afflict RHIC, especially because physicists relish challenging the ideas, calculations, and measurements of their colleagues.
  
The legal system, however, works differently.  One side almost invariably pushes its own point of view while dismissing or trying to demolish the other side's arguments.  If the opposition then cannot defend itself or convince a judge or jury, then it loses, even if it was ``right."  To a layperson or to a member of the legal profession, it thus makes good sense to have two opposing camps present their arguments for and against doing experiments at RHIC.  In fact, Dar and his colleagues actually pointed out loopholes in some of the arguments concerning the stability of the universe against nuclear collisions in RHIC, for example the argument of the Marburger committee that the moon has not been destroyed to date by cosmic rays:  A strangelet might be produced by the collision of an iron nucleus striking the surface of the moon, but immediately thereafter be fragmented by a collision with another nucleus just beneath the moon's surface, so that the strangelets would be destroyed before they have a chance to grow.  In general, therefore, should there not be two sides to this and other arguments, each pushing the limits of scientific knowledge and justifying its choice of uncertain parameters to produce an outcome most favorable to its position?

Posner cites the conclusion of Dar and his colleagues that the annual probability that strangelets produced at RHIC would destroy the Earth is less than 1 in 5 million, which seems like a very small number.  Consider, however, the reasoning of those who play the lottery.  A powerball jackpot of \$100 million, a ticket price of \$1, and 5 million tickets would be considered ``darn good odds," since the average return would be 20 to 1.  Who wouldn't play, given these odds?  Ignoring the obvious, this example points out the difference in what 1 in 5 million means in these two different contexts.  For the lottery, someone is eventually guaranteed to win.  For RHIC, there is no guarantee that RHIC would ever destroy the Earth no matter how long it runs.  The 1 in 5 million odds here may be interpreted as the maximum likelihood that the laws of Nature are such that strangelets are stable, that at least one would be created at RHIC, and that it would grow uncontrollably to destroy the Earth.  The odds are tiny but not zero.  A physicist never says never.  Is this tiny probability acceptable ... given the potentially devastating consequences?

Posner makes some excellent points regarding the relationship among science, society, and the legal system, as follows:
\begin{itemize}
\item ``The modern physical sciences concern phenomena of which ordinary people have no intuitive sense whatsoever, such as cell processes, the carbon cycle, and subatomic forces." \cite{27}
\item ``The legal system cannot deal effectively with scientifically and technologically difficult questions unless lawyers and judges -- not all, but more than at present -- are comfortable with such questions." \cite{28}
\end{itemize}

Posner's first point is well known to academic scientists, while his second is particularly provocative.  To address it he proposes that law schools should set up programs ``to assure that a nontrivial number of lawyers were comfortable with scientific methods, attitudes, usages, and vocabulary." \cite{29}  His word ``nontrivial" is a familiar one to physicists, and his idea is a splendid one, but many law schools may not be prepared to implement it.  Further, those few law students who have a background in science or engineering almost always seem to be attracted to patent law or closely related fields in which there is high demand and significant financial reward.  Posner, however, is advocating the education of scientifically knowledgeable lawyers who are excited about and willing to enter the judicial system and public service. I regard this as an entirely worthwhile goal that should be pursued by every concerned scientist and lawyer.

\section*{Conclusions}

I know of no professional physicist who is truly worried that a nuclear or particle-physics experiment could go so disastrously wrong that it could cause the end of the world.  Rather, to me the main lesson is that physicists must learn how to communicate their exciting discoveries to nonscientists honestly and seriously.  Perhaps Kenneth R. Foster put it best in his review of Posner's book: ``scientists must pay attention to the social as well as to the technical dimensions of technological risk--and develop a better understanding of how nonscientists will interpret their pronouncements on the subject." \cite{30}

I am a theorist who recently became a member of the Compact Muon Solenoid (CMS) experimental collaboration at the Large Hadron Collider (LHC ) at CERN.  This collaboration will carry out measurements on collisions of lead nuclei at center-of-momentum energies of 5.5 TeV (tera-electron volts; 1 TeV = $10^{12}$ eV) per nucleon energies.  This is greater than 1 PeV (peta-electron volt; 1 PeV = $10^{15}$ eV) total collision energy, which is the highest energy ever achieved in nuclear or particle-physics experiments.  We hope and expect that new physics beyond the Standard Model will be found.  All that I can safely predict with confidence, however, is that {\it the story I have related here will be repeated  at the LHC!}  In fact, a report entitled ``Study of Potentially Dangerous Events During Heavy Ion Collisions at the LHC" has already been prepared \cite{31}.  I do not believe that experiments at the LHC will cause the end of the world ... but I cannot prove it.

\section*{Appendix\footnote{I thank Arthur Poskanzer for sending me a copy of this report.}}

\noindent {\it Position Paper (Revised 5-14-79)}\\
 
\noindent Creation of Super-Dense Neutral Matter in the Bevalac\\

The possibility that stable super-dense matter can exist in nature, and might be formed during high-energy, heavy-ion collision experiments, was first suggested by Drs. Lee and Wick several years ago. The arguments advanced in support of this possibility were preliminary and inconclusive.  In 1974 in a panel discussion of this Lee-Wick matter at a conference at Bear Mountain [California] sponsored by the National Science Foundation, someone asked, ``If this matter could exist, what if it could also grow?" (i.e., by assimilating, or ``gobbling up" surrounding matter.) There was only a brief discussion and no conclusions were reached.

At a symposium at the Lawrence Berkeley Laboratory in November, 1977, Dr. A.S. Goldhaber of Stony Brook mentioned the discussion that had taken place at the Bear Mountain meeting.  Dr. Bernard Harvey, Associate Director for LBL's Nuclear Science Division, felt that the subject merited serious consideration.  He therefore recruited a select committee to discuss the subject. The committee met on the UC Berkeley campus in January 1978.

After 1$\frac{1}{2}$ days of discussion, the committee unanimously agreed that super-dense, super-stable neutral matter would not be formed in the Bevalac. This conclusion was based on both theoretical considerations and on pragmatic evidence.

Based upon present knowledge and theory, it was concluded that for such matter to be formed and to be capable of growth, it would have to be more stable than any known nuclear matter. In addition it would have to have no charge, i.e., be totally neutral.  If it were positively charged, like normal nuclear matter, it would repel other nuclei and so would be unable to grow by combining with them. Even if the required very high densities were achieved, it is very unlikely that the matter would also be stable and neutral.

If such an event could occur in a laboratory, it should also happen in the collision of cosmic rays on bodies in space, such as the moon.  The moon and other bodies in space are continually bombarded by particles at energies like those used in the Bevalac, yet, through billions of years, nothing of this kind has happened.

The committee agreed that collisions of heavy ions with nuclei at Bevalac energies should be studied, and that there was no need for special precautions to deal with the remote possibility of the formation of super-stable, neutral matter.

\indent The committee members were:\\
\indent Professor A. Kerman, MIT\\
\indent Professor Kinsey Anderson, Director, UCB Space Sciences Laboratory\\
\indent Professor A.S. Goldhaber, Stony Brook\\
\indent Dr. Miklos Gyulassy, Lawrence Berkeley Laboratory\\
\indent Professor T.D. Lee, Columbia University\\
\indent Professor Luis Alvarez, UCB and LBL, retired\\
\indent Dr. B.G. Harvey, Associate Director, Nuclear Science Division, LBL

\section*{Acknowledgments}

I thank Roger H. Stuewer for encouraging me to write this paper, and for his editorial work on it.  My research since I was a graduate student has been supported by the U.S. Department of Energy apart from one postdoctoral year by the U.S. National Science Foundation.

School of Physics and Astronomy\\
\indent University of Minnesota\\
\indent 116 Church Street SE\\
\indent Minneapolis, MN 55455 USA\\
\indent e-mail: kapusta@physics.umn.edu\\

\noindent
This paper was submitted to {\it Physics in Perspective} in January 2007 and is scheduled to appear in the June 2008 issue.

\end{document}